\documentstyle[12pt,twoside]{article}
\normalbaselines
\font\tbf = cmbx12

\pagebreak

\begin{document}

\indent
\vskip 1cm
\centerline{\tbf THE  BORN-INFELD  ELECTROMAGNETISM}
\vskip 0.3cm
\centerline{\tbf IN  KALUZA-KLEIN  THEORY}
\vskip 0.8cm
\centerline{by}
\vskip 0.3cm
\centerline{\tbf Jos\'e P. S. Lemos$^{(*)}$ and Richard Kerner}
\vskip 0.5cm
\centerline{\it Laboratoire de Gravitation et Cosmologie Relativistes}
\vskip 0.2cm
\centerline{\it Universit\'e Pierre et Marie Curie - CNRS ESA 7065}
\vskip 0.1cm
\centerline{\it Tour 22, 4$^{eme}$ \'etage, Bo\^{\i}te 142}
\vskip 0.1cm
\centerline{4, Place Jussieu, 75005 Paris, France}
\vskip 1.5cm
\indent
\hskip 0.5cm
{\tbf Abstract} 
\indent
{\small We investigate the properties of non-linear electromagnetism based
on the Born-Infeld Lagrangian in multi-dimensional theories of Kaluza-Klein
type. We consider flat space-time solutions only, which means that the
space-time metric is constant, and the only supplementary variable is the
dilaton field, a scalar. We show that in the case of Kaluza-Klein theory,
the Born-Infeld Lagrangian describes an interesting interaction between the
electromagnetic and scalar fields, whose propagation properties are 
modified in a non-trivial manner.}
\vskip 7.5cm
\indent
$(^*)$ : {\small Permanent address: Observat\'orio Nacional--Rio de Janeiro, 
also at IST--Lisbon.}

\newpage
\indent
{\tbf 1. Introduction}
\vskip 0.3cm
\indent
It is well known since the article by G. Boillat (\cite{Boillat}) that the 
Born-Infeld theory of electromagnetism (\cite{BI1,BI2,BI3}), although highly 
non-linear, leads under certain physical requirements to the propagation 
without birefringence, which makes this theory very special indeed. In 
addition, all types of disturbances and waves propagate without producing 
shocks in a finite time, implying that this case belongs to the class of 
theories called {\it completely exceptional}. 
\newline
\indent
The Born-Infeld Lagrangian is defined as the square root of the following 
determinant (we insist on using the mixed covariant and contravariant 
indeces, because only then the corresponding expression can be identified 
with a {\it matrix}, i.e. a linear operator, to which the concept of the 
determinant does apply): 
\begin{equation}
{\cal L} = \sqrt{{\rm det} ( {\delta_{\mu}}^{\nu} + {F_{\mu}}^\nu) } \, ,
			      \label{1.1} 
\end{equation}
where $F_{\mu\nu}=A_{\nu,\mu}-A_{\mu,\nu}$ is the Maxwell tensor
(partial derivatives in covariant indeces will be denoted by a comma).
This Lagrangian can be written also as ${\cal L}=\sqrt{1+2P-S^2}$ where
$P = \frac14 F_{\mu \nu} F^{\mu \nu}$ and $S = \frac14 ^*F^{\mu \nu}
F_{\mu \nu}$, with $^*F^{\mu\nu}=\frac12 \epsilon^{\mu\nu\rho\sigma}
F_{\rho \sigma}$ is the dual of $F_{\mu\nu}$.  In terms of the electric
and magnetic field intensities one has $P = \frac12 ({\bf B}^2 - {\bf
E}^2)$ and $S = {\bf E} \cdot {\bf B}$.  It has been proved in
\cite{Boillat} that the only other possible completely exceptional
theory comes from the singular Lagrangian ${\cal L} = P/S$.
\newline
\indent
It is amusing to note that the authors of this remarkable theory have
been led by the considerations of finiteness of energy, a natural postulate 
of recovering Maxwell's theory as linear approximation, and the hope to 
find soliton-like solutions representing charged particles like in the
non-linear (but non-relativistic) theory proposed by G. Mie (\cite{Mie}),  
rather than by an appropriate character of wave propagation \cite{Birula}.
\newline
\indent
Another important property that the Born-Infeld theory shares with
Maxwell's theory is its invariance under dual transformations, i.e.,
$SO(2)$ rotations of the fields $F_{\mu\nu}$ and $^*F^{\mu\nu}$ into
each other (see e.g. \cite{Deser}).
\newline
\indent
Originally, the Born-Infeld Lagrangian did not have any direct geometrical 
meaning. Another type of non-linear electrodynamics, with a clear geometrical 
rationale, which in the limit of weak field coincides with the Maxwell 
theory, but which differs from the Born-Infeld theory already in the second 
terms of its Taylor expansion, can be derived from the Kaluza-Klein theory 
in five dimensions, in the absence of scalar and gravitational fields.  
It is based on the fact that in five dimensions the Gauss-Bonnet term, 
$R_{A B C D} \, R^{A B C D} - 4 \, R_{A B } \, R^{A B} + R^2$ is not a 
topological invariant like in the $4$-dimensional case, but leads to 
non-trivial equations of motion of second order when added to the usual 
Einstein-Hilbert Lagrangian $R$. (The Gauss-Bonnet Lagrangian of the 
$5$-dimensional Kaluza-Klein theory including the variable scalar field has 
been derived by F. M\"uller-Hoissen in \cite{MullerHoissen}.) 
The full Lagrangian is \cite{Kerner1,Kerner2}:
\begin{equation}
{\cal L} = R + \gamma \,( R_{A B C D} \, R^{A B C D} 
- 4 \, R_{A B} \, R^{A B} + R^2 ) \, , \label{1.3} 
\end{equation}
with $\gamma$ being a dimensional parameter characterizing the relative 
weight of non-linear terms. When expressed in four dimensions in terms of 
the Maxwell tensor $F_{\mu \nu}$, one finds ${\cal L} = - \frac{1}{4} 
F_{\mu \nu}\, F^{\mu \nu} - \frac{3 \gamma}{16} \, ( F_{\mu \nu} F^{\mu \nu} 
)^2  - 2 \, (F_{\mu \lambda} F_{\nu \rho} F^{\mu \nu} F^{\lambda \rho}) , 
(\mu, \nu = 0,1,2,3).$ In terms of the invariants $P$ and $S$ this 
Lagrangian is given by ${\cal L}= 2P + \frac{3\gamma}{2}S^2$, which with an 
appropriate choice of $\gamma$ yields (up to a constant) the square of 
the Born-Infeld Lagrangian. This theory will be explored in section 5.
\newline
\indent
Recently, the Born-Infeld type Lagrangians appeared naturally in string
theories; this circumstance explains the revival of interest in the 
non-linear Born-Infeld electromagnetism and theories akin to it. For example, 
the interaction between the electromagnetic and scalar fields should be 
investigated because the dilaton (scalar) fields appear naturally in all 
effective string Lagrangians. Deser {\it et al.} \cite{Deser} have considered 
a theory based on the Lagrangian ${\cal L } = \sqrt{{\rm det}
({\delta_{\mu}}^{\nu} + {F_{\mu}}^{\ \nu} + \phi_{,\mu} \, 
\phi^{,\nu})}$ and have found that such theories do not ensure the 
absence of shock propagation. For the theory to be completely exceptional the 
total Lagrangian must be a sum of the Born-Infeld Lagrangian (or 
Maxwell's theory, or the singular $P/S$ theory mentioned above) and a
Lagrangian depending on the invariant $\phi_{,\mu} \, 
\phi^{,\mu})\,$ only, meaning the absence of any interaction
between those fields.
\newline
\indent
In spite of its very special properties, the Born-Infeld Lagrangian cannot 
be considered as the ultimate theory; the massless vector field it describes 
is certainly interacting with other possible excitations of string theories, 
in particular with the scalar field. We believe that one of the best ways 
of introducing the invariant interactions with the scalar field is the 
dimensional reduction technique. 
\newline
\indent
Instead of adding certain terms (out of several possible ones) containing 
the scalar field alone or interacting with the electromagnetic field, like 
in \cite{Deser}, we shall compute the Born-Infeld Lagrangian directly in 
$D+1+1$ dimensions (with $D=1,2,3$ being the number of space-like dimensions), which 
after reduction to $D+1$ spacetime dimensions will give us a unique 
expression containing the terms describing the contributions of the 
fields and their interactions.  We then study the propagation of dynamical 
fields in this setting.  Although the full theory is very highly non-linear 
and leads to very complicated characteristic cone equations, it is possible
to get some results in the case when one of the fields is constant (i.e.
does not propagate), but still can influence the propagation of the other 
one; partial results can be also obtained in the weak field limit.
\newline
\indent
Thus, we shall investigate systematically the combination of the Born-Infeld 
and Kaluza-Klein theories in lower dimensions, starting from the total 
dimension $3$, i.e. with one space and one time coordinate, plus one extra 
Kaluza-Klein dimension. We also analyze total dimensions $4$ and $5$.  
In general, the determinant of an arbitrary $N \times N$ matrix is a 
polynomial of $N$-th order containing all possible products of the elements 
of the matrix. However, we shall see that due to the particular structure 
of the matrix ${\delta_{\mu}}^{\nu} + {F_{\mu}}^{\ \nu}$, its determinants 
computed in odd dimensions $N = 2k + 1$ yield polynomials of even order $2k$ 
only.  
\vskip 0.3cm
\indent
{\tbf 2. Characteristic surfaces and field propagation}
\vskip 0.3cm
\indent
Our aim is to find out how do propagate the fields $F_{\mu \nu}$ and $\phi$, 
i.e. how the characteristic surfaces (cones) depend on dimension, whether 
both $F_{\mu \nu}$ and $\phi$ propagate along the same cones, and whether 
birefringence occurs or not, i.e., are these cones unique. The systems under 
consideration are second order partial differential equations, linear in the 
highest derivatives, with coefficients which depend only on the fields 
themselves. It is known (see e.g. \cite{Civita}) that the systems of this 
type can be reduced to a set of equations of first order in the derivatives 
via introduction of auxiliary fields. These auxiliary variables are the 
independent linear combinations of the first partial derivatives of functions 
describing the degrees of freedom of our system. For a scalar field these 
will be just all its first derivatives, $\partial_0 \phi \equiv \psi$, 
$\partial_i \phi \equiv \chi_i$ ($i$ representing the spatial indeces); 
for the electromagnetic potential $A_\mu$ the auxiliary variables are the 
electric field $E_i$ and magnetic field $B_i$ which are the only independent 
combinations of first order derivatives of $A_\mu$.
\newline
\indent
Next we represent the differential system by means of a matrix whose entries 
contain the operators of partial derivation or multiplicative coefficients,
acting on a vector-column representing auxiliary fields. If the vector-column  
${\bf u}$ with the fields $\psi$, $\chi_i$, $E_i$ and $B_i$ contains $N$
elements, then let us denote by ${\cal A}$ the $N$x$N$ matrix containing the 
partial derivatives and by ${\cal B}$ the $N$x$N$ matrix containing the
multiplicative factors. Then the field equations can be written in the 
form 
\begin{equation}
{\cal A}^{\mu} ({\bf u}) \partial_\mu {\bf u} + {\cal B}({\bf u})
{\bf u} = 0  \, . \label{2.1}
\end{equation} 
If the hypersurface defined by the implicit equation
\begin{equation}
\Sigma \, (x^{\mu})  \, = 0 \, .
                                      \label{2.2}
\end{equation} 
is a surface of discontinuity, then the first derivatives of fields are 
discontinuous across this surface, whereas the fields themselves are 
continuous. So, when applied to the {\it discontinuities} across the
hypersurface (\ref{2.2}), the equation (\ref{2.1}) reduces to 
\begin{equation}
\left({\cal A}^{\mu} \, \Sigma_{\mu} \right) \delta_1 {\bf u} = 0 \, , 
\label{2.3}
\end{equation}
where $\Sigma_{\mu} \equiv \partial_\mu \Sigma$, and $\delta_1 {\bf u}$
denotes the discontinuity of the first derivative across $\Sigma$,
$\delta_1 {\bf u} \equiv \partial {\bf u}/\partial\Sigma|_+ - \partial
{\bf u}/\partial\Sigma|_-$.  By definition, for a characteristic surface
one has $\delta_1 {\bf u}\ne 0$, therefore, in order for (\ref{2.3}) to
hold, one must have
\begin{equation} 
{\rm det} \biggl( {\cal A}^{\mu} \, \Sigma_{\mu} \biggr) = 0 \, , \label{2.4}
\end{equation}
on the surface of discontinuity. The characteristic equation (\ref{2.4}) 
determines the surface whose generic equation is $H(x,\Sigma_\mu)=0$, with 
$H$ a homogeneous function of order $N$ in $\Sigma_\mu$. The theories we 
study are {\it completely exceptional} since they obey the corresponding
condition of \cite{Boillat}, namely 
$\delta_0 H\equiv H|_+ - H|_- =0$ (see also \cite{Deser}).
\newline 
\indent
We will use this formalism to study the propagation of waves, characteristic 
equations and the possibility of birefringence in various theories. The 
solutions of the characteristic equation define the hypersurfaces along 
which the propagation takes place. If the solution is unique, it is said 
that there is no birefringence.
\newline
\indent
To make this setup clearer, let us show on a simple example how these
ideas work. We start with the simplest possible case: scalar field wave
equation in a two-dimensional space-time $(t,x)$:  
\begin{equation}
\partial^2_0 \, \phi - \partial^2_x \, \phi = 0 \, .
		  \label{2.5}
\end{equation}
(Partial derivatives in Lorentz indeces, $(0,x,y,z)$ or $(0,1,2,3)$,
will be denoted by $\partial$).  According to the prescription, we can
use as auxiliary fields $\psi$ and $\chi$ the first derivatives of the
scalar field $\phi$, $\partial_0 \phi = \psi $ and  $\partial_x \phi =
\chi$. Then by definition, the first derivatives of auxiliary fields
are not independent, because we have, as $\partial_0 \, (\partial_x \,
\phi) = \partial_x \, (\partial_0 \, \phi)$, automatically $\partial_0
\chi - \partial_x \psi = 0$.  On the other hand the dynamical equation
(\ref{2.5}) can be written as $\partial_0 \psi - \partial_x \chi = 0$.
In the matrix notation of (\ref{2.1}) these two equations can be combined
to yield 
\begin{equation}
\pmatrix{ 0 & 1 \cr 1 & 0} \, \partial_0  \pmatrix{ \psi \cr \chi} +
\pmatrix{ -1 & 0 \cr   0 & -1 } \, \partial_x \pmatrix{\psi \cr \chi}
= \pmatrix{ 0 \cr 0} \, .  
\label{2.6}
\end{equation}
We then find
\begin{equation}
{\cal A}^\mu\Sigma_\mu = \pmatrix{ - \Sigma_x & \Sigma_0 \cr 
\Sigma_0 & -\Sigma_x}\, ,
	    \label{2.7}
\end{equation}
and the characteristic equation ${\rm det}({\cal A}^\mu\Sigma_\mu)=0$ 
can be written as 
\begin{equation}
{\Sigma_0}^2 - {\Sigma_x}^2 = 0 \,  ,
	    \label{2.8}
\end{equation}
where $\Sigma_0\equiv\partial_0 \Sigma$ and $\Sigma_x\equiv\partial_x \Sigma$.
This last equation defines the characteristic surfaces $\Sigma(t,x)$,
which in this case are the light-cones in two space-time dimensions.

The same technique can be easily applied to the electromagnetic
Maxwellian field. The matrix becomes quite cumbersome, because we have
now six independent combinations of its first derivatives (the fields
${\bf E}$ and ${\bf B}$) appearing in the first-order equations of the
Maxwell system.  As we know, the characteristic surfaces in four
dimensions  are given by $\Sigma_{,\mu} \, \Sigma^{,\mu} = 
{\Sigma_0}^2 - {\Sigma_x}^2 -{\Sigma_y}^2 - {\Sigma_z}^2
=0$.  In the next sections we apply these techniques to more complex
cases in various dimensions.
\vskip 0.4cm
\indent
{\tbf 3. Born-Infeld type Lagrangians in two dimensions}
\vskip 0.3cm
\indent
{\tbf 3.1 The Born-Infeld theory}
\vskip 0.3cm
\indent
Starting from the lowest-dimensional case where there is only one space
and one time coordinate one finds that ${\rm det} ({\delta_{\mu}}^{\nu} 
+ {F_{\mu}}^\nu)$ has the form 
\begin{equation}
{\rm det} \, \pmatrix{ 1&F^0_{\ \ x} \cr F^x_{\ \ 0}& 1} 
= 1 - F^0_{\ \ x} \, F^x_{ \ \ 0}.
	    \label{3.1}
\end{equation}
Since  $F_{\mu\nu}=-F_{\nu\mu}$, the Born-Infeld Lagrangian can be 
written as
\begin{equation}
{\cal L} = \sqrt{ 1 + 2P},
	    \label{3.2}
\end{equation}
where
\begin{equation}
P\equiv\frac14 F_{\mu\nu}F^{\mu\nu},
	    \label{3.3}
\end{equation}
is the only invariant in two dimensions. The Euler-Lagrange equations 
are then 
\begin{equation}
{{\cal L}^{\alpha\beta}}_{,\alpha}=0, 
	    \label{3.4}
\end{equation}
with ${\cal L}^{\alpha\beta}={\cal L}_P F_{\alpha\beta}$ and  ${\cal L} = 
(\partial {\cal L}/ \partial P)$. The equation (\ref{3.4})
can be written as
\begin{equation}
(1+2P) \, {F^{\alpha\beta}}_{,\beta} - (F_{\mu\nu}F^{\mu\nu})_{,\beta} 
\, F^{\alpha\beta} = 0 \, . 
	    \label{3.5}
\end{equation}
In the linear approximation, when the second term of the above equation can 
be neglected, we get Maxwell's equations in two dimensions, i.e. 
${F^{\alpha\beta}}_{,\beta}=0$. The electric field is defined by 
$F^{0x}=\partial_x A_0 - \partial_0 A_x = E$, and the magnetic field does 
not exist. The equations for $E$ are:
\begin{equation}
\partial_0 E = 0 , \, \ \ \, \ \ \, \ \ \partial_x E = 0.
	    \label{3.6}
\end{equation}
We see that the only solution is $E={\rm const.}$ and there are no waves in 
Born-Infeld (nor in Maxwell) theory in a two-dimensional space-time.
\vskip 0.4cm
\indent
{\tbf 3.2 Born-Infeld theory in Kaluza-Klein space}
\vskip 0.3cm
\indent
With a compactified extra Kaluza-Klein dimension over the two-dimensional 
spacetime, and with constant metric tensor, we get the following expression
for  ${\rm det} ( {\delta_{\mu}}^{\nu} + {F_{\mu}}^\nu)$ 
containing the contribution of the scalar field:
\begin{eqnarray}
&
{\rm det} \, \pmatrix{ 
1 & F^0_{\ \ x} & \partial^0 \phi \cr F^x_{\ \ 0} & 1 & 
\partial^x \phi \cr \partial_0 \phi & \partial_x \phi & 1} = 
1 - F^0_{ \ \ x} 
\, F^x_{ \ \ 0} - \partial_0 \phi \partial^0 \phi - \partial_x \phi 
\partial^x \phi \, = 
&
\nonumber\\
&
= 1 + \frac12 F_{\mu\nu}F^{\mu\nu} - {\phi}_{,\mu} {\phi}^{,\mu} \, .
	    \label{3.7}
\end{eqnarray}
In the above expression one can note the absence of direct coupling between 
the scalar and gauge fields. Taking the square root of (\ref{3.7}) one 
finds the Born-Infeld Lagrangian :
\begin{equation}
{\cal L}=\sqrt{ 1 + 2P - 2\Phi}
	    \label{3.8}
\end{equation}
where $P$ is given by (\ref{3.3}) and  
\begin{equation}
\Phi\equiv\frac12 {\phi}_{,\mu}{\phi}^{,\mu} \, .
	    \label{3.9}
\end{equation}
The equations of motion for the vector potential $A_\mu$ and the scalar 
$\phi$ are, respectively
\begin{equation}
{\cal L}^2 \, {F^{\alpha\beta}}_{,\beta} - \frac14 
F^{\alpha\beta} \, (F_{\mu\nu}F^{\mu\nu})_{,\beta}
+\frac12 F^{\alpha\beta} \, (\phi^{,\mu}\phi_{,\mu})_{,\beta} = 0, 
	    \label{3.10}
\end{equation}
and 
\begin{equation}
{\cal L}^2  {\phi^{,\alpha}}_{,\alpha} - \frac14 \phi^{,\alpha} 
(F_{\mu\nu}F^{\mu\nu})_{,\alpha}
+\frac12 \phi^{,\alpha} (\phi^{,\mu}\phi_{,\mu})_{,\alpha} = 0.
	    \label{3.11}
\end{equation}
As before, we define the electric field $E$ as the derivative of  
the potential $A_\mu$
\begin{equation}
\partial_x A_0 - \partial_0 A_x = E \, ,
	    \label{3.12}
\end{equation}
and the two auxiliary fields $\psi$ and $\chi$ which are the first partial
derivatives of the field $\phi$,
\begin{equation}
\partial_0 \phi = \psi , \, \ \ \, \ \ \, \ \ \partial_x \phi = \chi \, .
	    \label{3.13}
\end{equation}
The equation (\ref{3.13}) implies
\begin{equation}
\partial_0 \chi - \partial_x \psi = 0 \, .
	    \label{3.14}
\end{equation}
Using (\ref{3.10}) and (\ref{3.12}) we obtain the equation of motion 
for the electric field 
\begin{equation}
(1-\psi^2+\chi^2) \, \partial_0 E + \psi \, E \, \partial_0 \psi -
\chi \, E \, \partial_0\chi = 0 \, ,
				   \label{3.15}
\end{equation}
and an analogous equation with $\partial_0$ replaced by $\partial_x$. 
From  (\ref{3.11}) and (\ref{3.13}) we get the equation of motion for 
the scalar field : 
\begin{equation}
(1+E^2+\chi^2) \, \partial_0 \psi - (1+E^2-\psi^2) \, \partial_x \chi - 
\psi \, \chi \, (\partial_x\psi+\partial_0 \chi) + \psi \, E \,\partial_0
E - \chi \, E \, \partial_x E = 0 
\, .
	    \label{3.16}
\end{equation}
Using the formalism introduced in section 2 we define the column-vector
$\bf u$ with the fields $E$, $\psi$ and $\chi$. Then the equations
(\ref{3.14}), (\ref{3.15}) and (\ref{3.16}) can be put in the following
matrix form:
\begin{eqnarray}
&\pmatrix{1-\psi^2+\chi^2 & \psi \, E & - \chi \, E
\cr
0 & 0 & 1
\cr 
\psi \, E & 1-E^2 + \chi^2 & - \psi \, \chi}
\partial_0
\pmatrix{E \cr \psi \cr \chi}
+
&
\nonumber\\
&+
\pmatrix{0 & 0 & 0
\cr
0 & -1 & 0
\cr 
- \chi \, E &  -\psi \, \chi & -(1-E^2-\psi^2) }
\partial_x
\pmatrix{E \cr \psi \cr \chi}
= \pmatrix{0 \cr 0 \cr 0}. &
	    \label{3.17}
\end{eqnarray}
>From the equations (\ref{3.17}), (\ref{2.1}) and (\ref{2.3}) we find 
\begin{eqnarray}
& {\cal A}^\mu \, \Sigma_\mu=
&
\nonumber\\
&
\pmatrix{(1-\psi^2 + \chi^2) \Sigma_0 & \psi E \Sigma_0 & - \chi E \Sigma_0
\cr
0 & -\Sigma_x & \Sigma_0
\cr 
\psi E \Sigma_0 - \chi E \Sigma_x & (1-E^2 + \chi^2) \Sigma_0
- \psi \chi \Sigma_x & - \psi \chi \Sigma_0 - (1-E^2-\psi^2) \Sigma_x}. &
	    \label{3.18}
\end{eqnarray}
Therefore the characteristic equation ${\rm det} ({\cal A}^\mu\Sigma_\mu)=0$ 
is : 
\begin{eqnarray}
&\Sigma_0 \left\{ \left[(1-\psi^2+\chi^2)(1-E^2+\chi^2)-
\psi^2 \, E^2 \right]{\Sigma_0}^2
\right.
&
\nonumber\\
&
-\left[(1-\psi^2+\chi^2)(1-E^2-\psi^2)+ \chi^2 \, E^2 \right]{\Sigma_x}^2
&
\nonumber\\
&
\left. - 2 \psi\chi \, \left[(1-\psi^2+\chi^2)-E^2\right]
\Sigma_0 \Sigma_x \right\} = 0 \, .
&
	    \label{3.19}
\end{eqnarray}
The equation (\ref{3.19}) can be written in the following covariant form:
\begin{eqnarray}
&
(V^\gamma\Sigma_\gamma)
\left[ (1-\phi_{,\beta}\phi^{,\beta}) (g^{\mu\nu} + 
F^{\mu\alpha}{F^{\nu}}_\alpha) \right.
&
\nonumber\\
&
\left.
-(1-\phi_{,\beta}\phi^{,\beta})
(\phi_{,\alpha}\phi^{,\alpha}g^{\mu\nu}-\phi^{,\mu}
\phi^{,\nu}) 
+ \frac12 F_{\alpha\beta}F^{\alpha\beta}\phi^{,\mu}\phi^{,\nu}
\right]\Sigma_\mu\Sigma_\nu = 0, &
	    \label{3.20}
\end{eqnarray}
with $V^\gamma$ a time-like vector. We see that the characteristic equation 
(\ref{3.20}) separates into a part that does not propagate the electric 
field and a part that propagates the scalar field.  Consequently, in a 
linear theory given by the Lagrangian (\ref{3.7}) (without the square root), 
the electric field $E$ does not propagate and the scalar field propagates
along the Minkowskian light-cones, $g^{\mu\nu}\Sigma_\mu\Sigma_\nu = 0$
or in Lorentzian coordinates $\Sigma_0^2-\Sigma_x^2=0$, independently of
the constant value of the field $E$.
\newline
\indent
To find the dispersion relation of the theory we put $\Sigma_\mu\equiv
\Sigma_{,\mu}$, and use the ansatz $\Sigma=a {\rm e}^{i(\omega t- kx)}$. 
Inserting into (\ref{3.19}) we find the relation $k=k(\omega)$ 
characterizing the propagation of the scalar field in this theory :
\begin{equation}
k=\pm\omega \frac{\sqrt{\beta^2+\alpha\gamma}\pm \beta}{\alpha} \, ,
	    \label{3.21}
\end{equation}
where $\alpha\equiv(1-\psi^2+\chi^2)(1-E^2-\psi^2)+E^2\chi^2$, 
$\beta=\psi\chi\left[(1-\psi^2+\chi^2)-E^2\right]$, and
$\gamma=(1-\psi^2+\chi^2)(1-E^2+\chi^2)-E^2\psi^2$.
To simplify this expression let us consider a scalar wave propagating in the
vacuum, $E=0$. Then (\ref{3.21}) reduces to 
\begin{equation}
k=\pm\omega \frac{\sqrt{1-\psi^2+\chi^2}\pm\psi\chi}{1-\psi^2}.
	    \label{3.22}
\end{equation}
In the weak field limit $\psi^2\sim \chi^2 \ll 1$ one has $k = \pm \omega 
\left[ 1 + \frac12 (\psi\pm\chi)^2 \right]$.  This means that the propagation 
velocity of the wave, $v\equiv\mid {\omega}/{k}\mid=1 / \left[ 1 + \frac12
(\psi\pm\chi)^2\right]$, is less than one, i.e., less than the speed of light 
in the linear theory.  We also note that (\ref{3.22}) gives no birefringence,  
as it should be expected for a a scalar field. There are only two solutions, 
one describing an outgoing, the other an ingoing wave.
\vskip 0.3cm
\indent
{\tbf 4. Born-Infeld type Lagrangians in 3 dimensions}
\vskip 0.3cm
{\tbf 4.1 Maxwell's theory}
\vskip 0.2cm
\indent
In three space-time dimensions Maxwell's theory is non-trivial. The 
Lagrangian is ${\cal L} = 2P$, where $P=\frac{1}{4} \, F_{\mu\nu}F^{\mu\nu}$. 
Introducing the four-potential $A_\mu$ we define the Maxwell tensor as 
before, 
\begin{equation}
F_{\mu\nu}= A_{\nu,\mu} - A_{\mu,\nu}. 
	    \label{4.1}
\end{equation}
\indent
In order to simplify the problem we consider here the characteristic
equation for a system depending only on two variables $t$ and $y$, and
in which the only non-vanishing fields are $A_x(t,y)$, $E_x(t,y)$ and
$B(t,y)$.  Despite of this simplified choices for the potentials (some
of the components suppressed, and depending only on some variables) the
result is covariant and unique.  For this type of simplifications see
e.g. \cite{Civita,couranthilbert}. From the equation (\ref{4.1}), and
defining $F^{0x}=E_x$, $F^{xy}=B$, we have $\partial_0 A_x + E_x = 0$
and $\partial_y A_x+B=0$. The two equations combined yield
\begin{equation}
\partial_0 B-\partial_y E_x =0 \, .
	    \label{4.3}
\end{equation}
The dynamical equation of motion is
\begin{equation}
\partial_y B-\partial_0 E_x =0.
	    \label{4.4}
\end{equation}
Then, using the formalism of section 2 we obtain 
\begin{equation}
{\cal A}^\mu\Sigma_\mu= \pmatrix{ -\Sigma_y & \Sigma_0  \cr 
-\Sigma_0 & \Sigma_y}  \, ,
	    \label{4.5}
\end{equation}
so the characteristic equation ${\rm det} \,({\cal A}^\mu\Sigma_\mu) = 0 \,$ becomes 
$$ \, \Sigma_0^2 - \Sigma_y^2=0 \, .$$  
Its invariant generalization taking into account the two space-like 
dimensions is obvious :
\begin{equation}
\Sigma_0^2 - \Sigma_x^2 - \Sigma_y^2=0 \, .
	    \label{4.6}
\end{equation}
Therefore, also in three space-time dimensions, Maxwell's theory leads to 
the propagation along the light-cones. 
\indent
\vskip 0.3cm
\indent
{\tbf 4.2 Maxwell's theory with non-interacting scalar field}
\vskip 0.2cm
\indent
Before we proceed to study Born-Infeld itself in 3D, and Born-Infeld with 
an extra Kaluza-Klein dimension let us study a non-interacting Maxwell and 
scalar field theory whose Lagrangian is given by
\begin{equation}
{\cal L} = 2P - 2\Phi \, ,
	    \label{4.7}
\end{equation}
where, as before, $P=\frac14 F_{\mu\nu}F^{\mu\nu}$ and $\Phi=\frac12 
\phi_{,\alpha}\, \phi^{,\alpha}$. Variation with respect to the 
electromagnetic potential $A_\rho$ yields Maxwell's equations 
${F^{\rho\sigma}}_{,\rho}=0$ whereas variation with respect to $\phi$ yields 
${\phi^{,\rho}}_{,\rho}=0$. In order to find the characteristic 
equations we again suppose that the system depends only on two variables 
$(t,y)$, with nonzero fields given by $A_x(t,y)$, $E_x(t,y)$, $B(t,y)$ and 
$\phi(t,y)$. Then with the definitons $\partial_0\phi - \psi=0$ and 
$\partial_y\phi-\chi=0$, we have the following set of equations,
\begin{eqnarray}
& - \partial_0 B + \partial_y E_x = 0 \, , \ \  \, \ \ \, 
- \partial_0 E_x + \partial_y B = 0 \, ,&
\nonumber\\
& \partial_0 \chi -\partial_y \psi= 0 \, , \ \ \, \ \ \,
\partial_0 \psi -\partial_y \chi= 0 \, . 
	    \label{4.8}
\end{eqnarray}
Calculating ${\cal A}^\mu\Sigma_\mu$ one obtains
\begin{equation}
{\cal A}^\mu\Sigma_\mu= \pmatrix{ \Sigma_y & -\Sigma_0 & 0 & 0 \cr 
	       -\Sigma_0 & \Sigma_y & 0 & 0 \cr
		       0 & 0 & -\Sigma_y & \Sigma_0 \cr
		       0 & 0 & \Sigma_0 & -\Sigma_y \cr} \, .
	    \label{4.9}
\end{equation}
The characteristic equation ${\rm det} ({\cal A}^\mu\Sigma_\mu) = 0$ is
$(\Sigma_0^2 -\Sigma_y^2)(\Sigma_0^2 -\Sigma_y^2)=0 \, . $
\newline
\indent
Again, the three-dimensional generalization is obvious:
\begin{equation}
(\Sigma_0^2-\Sigma_x^2-\Sigma_y^2)(\Sigma_0^2-\Sigma_x^2-\Sigma_y^2)=0 \, .
	    \label{4.10}
\end{equation}
We see that the characteristic is the equation for two non-interacting fields 
propagating with the velocity of light. The propagation of the two fields 
splits into two independent parts because the characteristic matrix 
(\ref{4.9}) is block-diagonal. 
\vskip 0.3cm
\newpage
{\tbf 4.3 The Born-Infeld theory}
\vskip 0.3cm
In a three-dimensional space-time the Born-Infeld determinant is :
\begin{eqnarray}
&
{\rm det} \, \pmatrix{1 & F^0_{ \ \ x} & F^0_{ \ \ y} \cr F^x_{ \ \ 0} 
& 1 &
F^x_{\ \ y} \cr F^y_{\ \ 0} & F^y_{ \ \ x} & 1} = 1 - F^0_{ \ \ x} \,
F^x_{\ \ 0} - F^0_{ \ \ y} \, F^y_{ \ \ 0} - 
F^x_{ \ \ y} \, F^y_{ \ \ x} =
&
\nonumber \\
&
= 1+ F_{\mu\nu}F^{\mu\nu} \, ,
	                             \label{4.11}
\end{eqnarray}
where we note the absence of terms of third order in the fields. Therefore, 
taking the square root, we get the same action as in two dimensions, i.e. 
${\cal L} = \sqrt{ 1 + 2P}$ but now $F^{0x}=E_x$, $F^{0y}=E_y$ and $F^{xy}=B$ 
as in Maxwell's theory.  The dynamical equation of motion, given by 
${{\cal L}^{\alpha\beta}}_{,\alpha}=0$ with ${\cal L}^{\alpha\beta} 
= {\cal L}_P F^{\alpha\beta}$ yields                                                     
\begin{equation}
(1+2P) {F^{\alpha\beta}}_{,\alpha} - \frac14 
(F_{\mu\nu}F^{\mu\nu})_{,\alpha}F^{\alpha\beta}=0 \, .  
	    \label{4.12}
\end{equation}
Combining (\ref{4.1}) and (\ref{4.12}), we get the following set of six 
equations,
\begin{eqnarray}
& \partial_x A_0 -\partial_0 A_x -E_x = 0 \, , &
\nonumber\\
&  \partial_y A_0 -\partial_0 A_y -E_y = 0 \, ,&
\nonumber\\
& \partial_x A_y -\partial_y A_x - B= 0 \, ,&
\nonumber\\
&
(1+B^2-E_y^2)\partial_x E_x + (1+B^2-E_x^2) \partial_y E_y +&
\nonumber\\
& E_x E_y \partial_x E_y
-E_x B \partial_x B + E_x E_y \partial_y E_x - E_y B \partial_y B = 0 
\, , &
\nonumber\\
&
(1-E_x^2-E_y^2)\partial_y B - (1+B^2-E_y^2) \partial_0 E_x - 
&
\nonumber\\
& E_x E_y \partial_0 E_y
+E_x B \partial_0 B + E_x B \partial_y E_x + E_y B \partial_y E_y = 0 
\, , &
\nonumber\\
&
(1-E_x^2-E_y^2)\partial_x B + (1+B^2-E_x^2) \partial_0 E_y + 
&
\nonumber\\
& E_x E_y \partial_0 E_x
-E_y B \partial_0 B + E_x B \partial_x E_x + E_y B \partial_x E_y = 0 
\, .
	    \label{4.13}
\end{eqnarray} 
As we did in the previous section, where we have chosen a reduced system, we shall
study the characteristics for a system depending only on two variables 
$(t,y)$ and with nonzero fields given by $A_x(t,y)$, $E_x(t,y)$ and $B(t,y)$. 
In this case in (\ref{4.13}) only two independent equations remain :
\begin{eqnarray}
& \partial_0 B-\partial_y E_x =0 \, ,
&
\nonumber\\
& (1-E_x^2)\partial_y B - (1+B^2) \partial_0 E_x +E_x B \partial_0 B 
+ E_x B \partial_y E_x  = 0 \, . &
	    \label{4.14}
\end{eqnarray}
Using the formalism of section 2 we obtain
\begin{equation}
{\cal A}^\mu\Sigma_\mu= \pmatrix{ -\Sigma_y & \Sigma_0  \cr 
-(1+B^2)\Sigma_0 + E_xB\Sigma_y &  E_xB\Sigma_0 + (1+E_x^2)\Sigma_y} \, .
	    \label{4.15}
\end{equation}
Thus the characteristic equation 
${\rm det}({\cal A}^\mu\Sigma_\mu)=0$ can be put in the form,
\begin{equation}
\Sigma_0^2 -\Sigma_y^2 + (B\Sigma_0-E_x\Sigma_y)^2 =0 \, .
	    \label{4.16}
\end{equation}
In 3D the dual to $F_{\alpha\beta}$ is 
\begin{equation}
^*{F^\gamma} = \frac{1}{2} \, \epsilon^{\alpha \beta \gamma}F_{\alpha \beta}, 
	    \label{4.2}
\end{equation}
where $\epsilon^{\alpha \beta \gamma}=\frac{1}{\sqrt{-g}} 
e^{\alpha\beta\gamma}$ is the Levi-Civit\`a tensor and 
$e^{\alpha\beta\gamma}$ the Levi-Civit\`a symbol. 
\newline
\indent
One can write the characteristic equation (\ref{4.16}) in a covariant 
form, 
\begin{equation}
(g^{\alpha\beta} + {^*{F^\alpha}} \, {^*{F^\beta}}) 
\Sigma_\alpha \Sigma_\beta = 0    
 \, .
	    \label{4.17}
\end{equation}
Since $\Sigma_\mu\equiv \Sigma_{,\mu}$, for some function $\Sigma$, one 
can seek the ansatz $\Sigma=a {\rm e}^{i(\omega t- k y)}$ to find with 
the help of (\ref{4.16}) the covariant expression
\begin{equation}
k=\pm\omega \frac{\sqrt{1+B^2-E^2}\mp E B}{1-E^2}\, ,
	    \label{4.18}
\end{equation}
with $E = \sqrt{E_x^2 + E_y^2}.$ This means that the propagation velocity 
is less than the speed of light in the linear theory, and there is no 
birefringence as it is expected for a Born-Infeld theory, \cite{Boillat}.
\vskip 0.3cm
\indent
{\tbf 4.4 Born-Infeld theory in Kaluza-Klein space}
\vskip 0.2cm
Adding the extra Kaluza-Klein dimension gives:
\begin{eqnarray}
&
{\rm det}\, \pmatrix{
1 & F^0_{ \ \ x} & F^0_{ \ \ y} & \partial^0 
\phi \cr 
F^x_{ \ \ 0} & 1 & F^x_{\ \ y} & \partial^x 
\phi \cr F^y_{ \ \ 0} & 
F^y_{\ \ x} & 1 & \partial^y \phi \cr 
\partial_0 \phi & \partial_x \phi & 
\partial_y \phi & 1} = 
&
\nonumber\\
&
1 + 
\frac12 F_{\mu\nu}F^{\mu\nu} - {\phi}_{,\mu} {\phi}^{,\mu} - 
{\phi}_{,\mu} {\phi}_{,\nu}
(\frac12
F_{\alpha\beta}F^{\alpha\beta}g^{\mu\nu}-
F^{\mu\alpha}{F^\nu}_\alpha) \, .
	    \label{4.19}
\end{eqnarray}
The Born-Infeld type Lagrangian is then
\begin{equation}
{\cal L}=\sqrt{ 1 + 2P - 2\Phi -2I} \, ,
	    \label{4.20}
\end{equation}
where $P=\frac14 F_{\mu\nu}F^{\mu\nu}$, 
$\Phi=\frac12 {\phi}_{,\mu} {\phi}^{,\mu}$ and 
$I = \frac12 {\phi}_{,\mu}{\phi}_{,\nu} (\frac12
F_{\alpha\beta}F^{\alpha\beta}g^{\mu\nu}-F^{\mu\alpha}{F^\nu}_\alpha)$. 
The variation with respect to the vector potential $A_\sigma$ yields
\begin{eqnarray}
& {\cal L}^2 \left[1-({\phi}_{,\mu}{\phi}^{,\mu})\right]
{F^{\alpha\beta}}_{,\alpha}
-\left[{\cal L}{\cal L}_{,\alpha} (1-{\phi}_{,\mu} {\phi}^{,\mu})
+{\cal L}^2({\phi}_{,\mu}{\phi}^{,\mu})_{,\alpha}
F^{\alpha\beta}\right] + &
\nonumber\\
& 
2 {\cal L}^2  {\phi}^{,\alpha} {\phi}_{,\lambda} 
{F^{\lambda\beta}}_{,\alpha}
+ &
\nonumber\\
&
2\left[ {\cal L}^2 {\phi}^{,\alpha} {\phi}_{,\lambda,\alpha} +
{\cal L}^2  {{\phi}^{,\alpha}}_{,\alpha} {\phi}_{,\lambda} -
{\cal L}{\cal L}_{,\alpha}{\phi}^{,\alpha} {\phi}_{,\lambda}
\right] F^{\lambda\beta}=0 \, . &
                        \label{4.21}
\end{eqnarray}
and the variation with respect to the scalar field $\phi$ gives
\begin{eqnarray}
& {\cal L}^2 (1+\frac12 F_{\mu\nu}F^{\mu\nu}){\phi^{,\alpha}}_{,\alpha}
- {\cal L}{\cal L}_{,\alpha} \phi^{,\alpha}  + {\cal L}^2\phi^{,\alpha}
(\frac12 F_{\mu\nu}F^{\mu\nu})_{,\alpha} 
 &
\nonumber\\
& - {\cal L}^2 \left[ 
(F^{\alpha\sigma}{F^{\lambda}}_\sigma)_{,\alpha}\phi_{,\lambda}
+F^{\alpha\sigma}{F^{\lambda}}_\sigma\phi_{,\lambda,\alpha}
\right] 
&
\nonumber\\
&
-{\cal L}{\cal L}_{,\alpha} (\frac12 F_{\mu\nu}F^{\mu\nu}\phi^{,\alpha}
-F^{\alpha\sigma}{F^{\lambda}}_\sigma\phi_{,\lambda})
=0 \, . 
                        \label{4.22}
\end{eqnarray}
\indent
Again, we choose a reduced system to study the characteristic equations 
for the case depending only on two variables $(t,y)$ and with nonzero 
fields given by $A_x(t,y)$, $E_x(t,y)$ $B(t,y)$, and $\phi(t,y)$, with 
$E_x = -\partial_0 A_x$, $B = - \partial_y A_x$, $\psi = \partial_0\phi$ 
and $\chi = \partial_y\phi$.  The invariants are now $P=\frac12 (B^2-E^2)$, 
$\Phi=\frac12(\psi^2-\chi^2)$ and $I=\frac12(E\chi-B\psi)^2$.
\newline 
\indent
Now we can develop the formalism of section 2, but the high degree of 
non-linearity makes the problem almost intractable. The defining
equations are still the same, very simple. The complete set of 
four equations is 
\begin{eqnarray}
& -\partial_0 B + \partial_y E = 0 \, ,&  \nonumber
\\
& {\cal L}^2\left[1-(\psi^2-\chi^2)\right]
(\partial_0{E_x} - \partial_yB)&\nonumber\\& 
-E_x\left[ {\cal L}\partial_0{\cal L}(1-(\psi^2-\chi^2)) +
            {\cal L}^2 \partial_0(\psi^2-\chi^2)\right]&\nonumber\\& 
+B\left[ {\cal L}\partial_y{\cal L}(1-(\psi^2-\chi^2)) +
            {\cal L}^2\partial_y(\psi^2-\chi^2)\right]&\nonumber\\& 
+2{\cal L}^2\left(\partial_0\psi^2{E_x} - \psi\chi \partial_0 B - 
                                         \psi\chi\partial_y{E_x} +
                            \chi^2\partial_yB\right)&\nonumber\\& 
+2E_x\left[ {\cal L}^2(2\psi\partial_0\psi - \chi \partial_y\psi - 
                                      \psi \partial_y\chi) -
                {\cal L}\partial_0{\cal L}\psi^2 + 
                {\cal {\cal L}}\partial_y{\cal {\cal L}}
                 \psi\chi\right]&\nonumber\\& 
-2B\left[   2{\cal L}^2(-2\chi\partial_0\chi + 
 \psi \partial_0\chi + \chi \partial_0\psi) -
                 {\cal L}\partial_0{\cal L}\psi\chi + 
                 {\cal L}\partial_y{\cal  L} 
                 \chi^2\right] = 0 \, ,& \nonumber
\\
& \partial_0\chi-\partial_y \psi=0 \, ,& \nonumber
\\
&  -{\cal L}^2\left(1+(B^2-E^2)\right) 
      (\partial_0\psi-\partial_y\chi)+ &\nonumber\\& 
     \psi {\cal L}\partial_0{\cal L}-\chi {\cal L}\partial_y{\cal L} - 
      {\cal L}^2\psi\partial_0(B^2-E^2) + 
       {\cal L}^2 \partial_y\chi(B^2-E^2)
                                                            &\nonumber\\&
+{\cal L}^2\left[ -\partial_0({E_x}^2) 
    \psi + \partial_0(E_xB)\chi - \partial_y(E_xB)\psi \right.
                                                            &\nonumber\\&
      \left. - \partial_y(B^2)\chi -{E_x}^2 \partial_0\psi 
            + E_xB\partial_0\chi
      + E_xB\partial_y\psi - B^2 \partial_y\chi \right]&\nonumber\\&
+ {\cal L}\partial_0{\cal L}\left[ \psi(B^2-E^2) 
-(-{E_x}^2\psi+E_xB\chi)\right]&\nonumber\\&
+ {\cal L}\partial_y{\cal L}
   \left[ -\chi(B^2-E^2) -({E_x}B\psi-B^2\chi)\right] = 0
\, . 
                                 \label{4.23}
\end{eqnarray}
Then ${\cal A}^\mu\Sigma_\mu$ has the following structure, 
\begin{equation}
{\cal A}^\mu\Sigma_\mu= \pmatrix{ \Sigma_y & -\Sigma_0 & 0 & 0 \cr a_1\Sigma_0 
- a_5 \Sigma_y & a_2\Sigma_0-a_6\Sigma_y & a_3\Sigma_0-a_7 \Sigma_y 
& -a_4\Sigma_0 +a_8\Sigma_y \cr 0 & 0 & -\Sigma_y & \Sigma_0 \cr
b_1\Sigma_0 -b_5\Sigma_y & -b_2\Sigma_0+b_6\Sigma_y & b_3\Sigma_0 
-b_7\Sigma_y & -b_4\Sigma_0-b_8\Sigma_y \cr} \, ,
	    \label{4.24}
\end{equation}
where the coefficients $a$'s and $b$'s are known functions of $E$, $B$, 
$\psi$ and $\chi$. We can see how, due to the non-linearity, the off-diagonal 
terms start to appear signaling the interaction between the electromagnetic
and scalar fields. Now ${\rm det}({\cal A}^\mu\Sigma_\mu) = 0$ yields the
following characteristic equation which we put in covariant notation
\begin{equation}
G^{\alpha\beta\gamma\delta}
\Sigma_\alpha\Sigma_\beta\Sigma_\gamma\Sigma_\delta=0 \, ,
	    \label{4.25}
\end{equation}
with $G^{\alpha\beta\gamma\delta}={g_1}^{\alpha\beta}{f_1}^{\gamma\delta}
+{g_2}^{\alpha\beta}{f_2}^{\gamma\delta}$, where the tensor 
${g_1}^{\alpha\beta}$ is given by
\begin{eqnarray}
& {g_1}^{\alpha\beta} = 
(\eta^{\alpha\beta} + {^*{F^\alpha}} {^*{F^\beta}} 
- \phi_{,\mu}\phi^{,\mu} \eta^{\alpha\beta})
+&
\nonumber\\
& 
\left[ (2\phi^{,\alpha}\phi^{,\beta}-
\phi_{,\mu}\phi^{,\mu} \eta^{\alpha\beta})(1+2Q-2I)
-2P\phi^{,\alpha}\phi^{,\beta}+
+2I\eta^{\alpha\beta}
+{^*{F^\alpha}}\phi^{,\beta}\sqrt{2I}\right] 
&
\nonumber\\
& 
+\sqrt{2I}\sqrt{2\Phi}{^*{F^\alpha}}\phi^{,\beta}
\, , 
	    \label{4.26}
\end{eqnarray}
the tensor ${g_2}^{\alpha\beta}$ is given by
\begin{eqnarray}
& 
{g_2}^{\alpha\beta} = - \{ F^\alpha \phi^{,\beta} + 
\left[ -\phi_{,\mu}\phi^{,\mu}F^\alpha \phi^{,\beta}
+2 F^\mu \phi_{,\mu}(\phi^{,\alpha}\phi^{,\beta}-
\phi_{,\mu}\phi^{,\mu}\eta^{\alpha\beta}) \right]
&
\nonumber\\
& 
+\left[ 2 F^\mu \phi_{,\mu} (1+2P) - \sqrt{2I}\frac12 {^*{F^\mu}}F_\mu\right]
   \eta^{\alpha\beta} 
&
\nonumber\\
& \left.
-\left[ 2(2I)F^\alpha \phi^{,\beta} + \sqrt{2I}{^*{F^\mu}}\phi_{,\mu}
    (\phi_{,\nu}\phi^{,\nu}) \eta^{\alpha\beta} \right]
\right\}
\, , 
	    \label{4.27}
\end{eqnarray}
${f_1}^{\gamma\delta}$ is
\begin{eqnarray}
& {f_1}^{\gamma\delta} = 
(1+2P-2I)\eta^{\gamma\delta} - 
(\phi_{,\mu}\phi^{,\mu}\eta^{\gamma\delta}-\phi^{,\gamma}\phi^{,\delta})
&
\nonumber\\
& 
+(1+2P-2\Phi){^*{F^\gamma}}{^*{F^\delta}}
-2\sqrt{2I}{^*{F^\gamma}}\phi^{,\delta}
\, , 
	    \label{4.28}
\end{eqnarray}
and ${f_2}^{\gamma\delta}$ is
\begin{eqnarray}
& {f_2}^{\gamma\delta} = 
F^\gamma\phi^{\delta} + F^\mu\phi_{,\mu}
(\phi^{,\gamma}\phi^{,\delta}
-\phi_{,\mu}\phi^{,\mu}\eta^{\gamma\delta})
&
\nonumber\\
& 
\left[F^\mu\phi_{,\mu} \eta^{\gamma\delta} - F^\gamma\phi^{\delta}
+ F^\mu\phi_{,\mu}{^*{F^\gamma}}{^*{F^\delta}}\right]
\, .
	    \label{4.29}
\end{eqnarray}
Here $F^\mu=(E,0,B)$ and ${^*{F^\mu}}=(B,0,-E)$, this last as defined above.
\newline
\indent
One can in principle find solutions to equation (\ref{4.25}) under simplified 
assumptions, such as considering one field a constant while the other is 
propagating. We will do this for the four-dimensional case in section 5.4.
\vskip 0.4cm
{\tbf 5. Born-Infeld type Lagrangians in 4 dimensions}
\vskip 0.3cm
\indent
Now we consider the following theories in four dimensions: Maxwell's theory, 
Kaluza-Klein theory with Gauss-Bonnet term, the usual Born-Infeld electrodynamics,  
and the Born-Infeld theory in Kaluza-Klein space. 
\vskip 0.3cm
{\tbf 5.1 Maxwell's theory}
\vskip 0.3cm
\indent
The results for Maxwell's theory are well known. We choose a reduced system 
in order to provide a short demonstration. It is enough to consider the 
characteristic equations for a system depending only on two variables $(t,z)$ 
and with nonzero fields given by $A_x(t,z)$, $A_y(t,z)$, $E_x(t,z)$, 
$E_y(t,z)$, $B_x(t,z)$ and $B_y(t,z)$, with $E_x=-\partial_0 A_x$, 
$E_y=-\partial_0 A_y$, $B_x=-\partial_z A_y$ and $B_y=\partial_z A_x$. 
We obtain the following set of four equations, 
\begin{eqnarray}
& 
\partial_0 B_y + \partial_z E_x = 0 \, ,
&
\nonumber\\
& 
-\partial_0 B_x + \partial_z E_y = 0 \, , 
&
\nonumber\\
& 
\partial_z B_x - \partial_0 E_y = 0 \, ,
&
\nonumber\\
& 
\partial_z B_y + \partial_0 E_x = 0 \, . &
	    \label{5.1}
\end{eqnarray}
Then using the column-vector ${\bf u}=(E_x,E_y,B_x,B_y)$ one finds that the 
characteristic equation is
${\rm det}({\cal A}^\mu\Sigma_\mu)=({\Sigma_0}^2-{\Sigma_z}^2)^2=0$, or in
covariant form, $(g^{\mu\nu}\Sigma_\mu \Sigma_\nu)^2=0$. Thus there is 
no birefringence in Maxwell's theory, a well known fact.
\vskip 0.3cm
\newpage
{\tbf 5.2 The Gauss-Bonnet Lagrangian in Kaluza-Klein theory}
\vskip 0.3cm
As mentioned in the Introduction one may consider another non-linear
generalization of electrodynamics that can be derived from the
Kaluza-Klein theory in five dimensions, and is based on the addition of
the Gauss-Bonnet term, $R_{A B C D} \, R^{A B C D} - 4 \, R_{A B } \,
R^{A B} + R^2$, which in five dimensions is no more a topological
invariant, but leads to non-trivial equations of motion of second order
when added to the usual Einstein-Hilbert Lagrangian $R$.
\newline
\indent
In a flat space-time and without the scalar field the Kaluza-Klein metric is
\begin{equation}
g_{AB} = \pmatrix{ g_{\mu \nu} + A_{\mu} A_{\nu} 
& A_{\mu} \cr A_{\nu} & 1} \, ,
			      \label{5.2} 
\end{equation}
where $A,B=0,1,2,3,5$ and $\mu,\nu=0,1,2,3$ (or $\mu,\nu=0,x,y,z$ following 
the covention we have been using). 
As mentioned the full Lagrangian is taken to be (see 
\cite{Kerner1,Kerner2}):
\begin{equation}
{\cal L} = R + \gamma \,( R_{A B C D} \, R^{A B C D} 
- 4 \, R_{A B} \, R^{A B} + R^2 ) \, ,
			      \label{5.3} 
\end{equation}
with $\gamma$ being a certain dimensional parameter characterizing the
strength of the non-linearity. When expressed in four dimensions in terms 
of the Maxwell tensor, it becomes 
\begin{equation}
{\cal L} = - \frac{1}{4} F_{\mu \nu}\, F^{\mu \nu} - \frac{3 \gamma}{16}
\, ( F_{\mu \nu} F^{\mu \nu} )^2  - 2 \, (F_{\mu \lambda} F_{\nu \rho}
F^{\mu \nu} F^{\lambda \rho}) \, .
			     \label{5.4}
\end{equation}
In terms of the invariants $P$ and $S$ this Lagrangian is given by ${\cal L}
= 2P + \frac{3\gamma}{2} S^2$, which for the choice $\gamma=-\frac{2}{3}$
yields essentially the square of the Born-Infeld Lagrangian. 
The equations of motion are:
\begin{equation}
F_{\lambda \rho,\mu} +  F_{\rho \mu,\lambda} +
F_{\mu \lambda,\rho} = 0 \, ,
			     \label{5.5}
\end{equation}
which correspond to the Bianchi identities and are geometrical equations 
valid independently of the Lagrangian chosen, and the dynamical equations 
resulting from the variational principle,
\begin{equation}
[ F^{\lambda \rho} - \frac{3 \gamma}{2} \, 
(F_{\mu \nu}
F^{\mu \nu}) F^{\lambda \rho} + 
\frac{3 \gamma}{2} F_{\mu \nu} F^{\lambda \mu}
F^{\rho \nu} ]_{,\lambda}= 0 \, . 
			     \label{5.6}
\end{equation}
The Lagrangian (\ref{5.4}) is particularly simple when expressed in more 
familiar terms with the fields $\bf E$ and $\bf B$ :
\begin{equation}
{\cal L } = \frac{1}{2}( {\bf B}^2 - {\bf E}^2) + 
\frac{3 \gamma}{2} \,
({\bf E} \cdot {\bf B} )^2
			     \label{5.7}
\end{equation}
The equations of motion also display a clear physical meaning when 
expressed in terms of $\bf E$ and $\bf B$. The equation (\ref{5.6}) becomes
\begin{equation}
{\bf div}\,{\bf B} = 0 , \, \ \ \, \ \ \, \ \ {\bf rot} \, {\bf E} = - 
\partial_0 {\bf B} ,
			     \label{5.8}
\end{equation}
whereas the equations (\ref{5.7}) become
\begin{eqnarray}
& {\bf div} \, {\bf E} = - 3 \gamma \, {\bf B} 
\cdot {\bf grad} \, ( {\bf E} \cdot {\bf B}) 
&\nonumber\\&
{\bf rot} \, {\bf B} = \partial_0 {\bf E} + 3 \, \gamma \, 
\biggl[ \, {\bf B} \partial_0 ({\bf E} \cdot {\bf B}) 
- {\bf E} \times {\bf grad} ( {\bf E} \cdot {\bf B}) \biggr] \, .
			     \label{5.9}
\end{eqnarray}
which show how the density of charge and the current are created by the
non-linearity of the field: indeed, we can introduce
\begin{equation}
\rho = - 3 \gamma \, {\bf B} \cdot {\bf grad} \, ( {\bf E} \cdot 
{\bf B}) \, \ \ \, \ \ {\rm and} \, \ \ \, \ \ \, {\bf j} =
3 \, \gamma \, \biggl[ \, {\bf B} 
\partial_0 ({\bf E} \cdot {\bf B})
- {\bf E} \times {\bf grad} ( {\bf E} \cdot {\bf B}) \biggr] 
			     \label{5.10}
\end{equation}
which satisfy the continuity equation
\begin{equation}
\partial_0 \rho + {\bf div} \, {\bf j} = 0
\end{equation}
The Poynting vector conserves its form known from the Maxwellian theory, but
the energy density is modified:
\begin{equation}
{\bf S} = {\bf E} \times {\bf B} , \, \ \ \, \ \ \, {\cal{E}} = 
\frac{1}{2} \,
({\bf E}^2 + {\bf B}^2) + 
\frac{3 \gamma}{2} \, ({\bf E} \cdot {\bf B})^2 \, ,
			    \label{5.11}
\end{equation}
with the continuity equation resuming the energy conservation satisfied by
virtue of the equations of motion:
\begin{equation}
\partial_0 {\cal E} + {\bf div} {\bf S} = 0 \, .
			    \label{5.12}
\end{equation}
The properties of possible stationary axisymmetric solutions, endowed
with non-vanishing charge, intrinsic kinetic and magnetic moments, have
been discussed in \cite{Kerner1,Kerner2}. In the theory based on the 
Gauss-Bonnet term in $5$ dimensions, whose main features were developed in 
section 1, the first two equations of (\ref{5.1}) are the same here, but the 
dynamical equations are instead given by
\begin{eqnarray}
& 
(1-{E_x}^2)\partial_z B_x - (1+{B_y}^2)\partial_0 E_y 
-B_yB_x\partial_0E_x-E_xB_y\partial_0B_x -E_yB_y\partial_0B_y&\nonumber\\& 
-E_xB_x\partial_zE_x-E_xE_y\partial_zB_y-E_xB_y\partial_zE_y=0 \, ,
&
\nonumber\\
& 
(1-{E_y}^2)\partial_z B_y + (1+{B_x}^2)\partial_0 E_x 
+E_xB_x\partial_0B_x+B_xB_y\partial_0E_y +E_yB_x\partial_0B_y&\nonumber\\& 
-E_yB_x\partial_zE_x-E_xE_y\partial_zB_y-E_yB_y\partial_zE_y=0 \, . 
	    \label{5.13}
\end{eqnarray}
Using again the column-vector ${\bf u}=(E_x,E_y,B_x,B_y)$ one finds that
the characteristic equation is now
\begin{eqnarray}
& 
{\rm det}({\cal A}^\mu\Sigma_\mu)=\left( {\Sigma_0}^2-{\Sigma_z}^2 \right)
                                                                  \times 
&
\nonumber\\
& 
\left[({\Sigma_0}^2-{\Sigma_z}^2)+({B_x}^2+{B_y}^2){\Sigma_0}^2
+2(E_xB_y-E_yB_x)\Sigma_0\Sigma_z + ({E_x}^2+{E_y}^2){\Sigma_z}^2\right] 
&
\nonumber\\
& 
=0\, .
	    \label{5.14}
\end{eqnarray}
This can be written in covariant form in a compact manner
\begin{equation}
\left(    g^{\mu\nu} \Sigma_\mu\Sigma_\nu \right)
\left[
(g^{\mu\nu}-{^*{{F^\mu}}_\alpha}{^*{F^{\mu\alpha}}})
\Sigma_\mu\Sigma_\nu   \right]=0 \, ,
	    \label{5.15}
\end{equation}
where ${^*{F^{\alpha\beta}}}\equiv\frac12
\epsilon^{\alpha\beta\gamma\delta}F_{\gamma\delta}$ and
$\epsilon^{\alpha\beta\gamma\delta}$ is the Levi-Civita tensor. Thus from 
(\ref{5.3}) we can see that there is birefringence. One wave propagates in
a Maxwellian way, the other propagates differently, in fact, it is delayed.
\newline 
\indent
Indeed, if we set that the wavefront has the form $\Sigma =a {\rm e}^{i(\omega t-kz)}$, 
we find two solutions given by
\begin{equation}
k=\pm \omega \, ,
	    \label{5.16}
\end{equation}
and
\begin{equation}
k=\pm \omega \frac{
\sqrt{(E\times B)^2+(1+B^2)(1-E^2)}\mp(E\times B)
}{1-E^2}\, .
	    \label{5.17}
\end{equation}
For (\ref{5.17}) one has that the wave velocity  $\omega/k$ is in general 
less than one, and it lags behind the other wave. 
\vskip 0.3cm
{\tbf 5.3 The Born-Infeld theory}
\vskip 0.3cm
\indent
Now, in four dimensions there is a new invariant, $S$ entering the 
Born-Infeld action, which is
\begin{equation}
{\cal L}=\sqrt{1+2P-S^2} \, ,
	    \label{5.18}
\end{equation}
where, $P=\frac14 F_{\mu\nu}F^{\mu\nu}$ and $S=\frac14 {^*{F_{\mu\nu}}}
F^{\mu\nu}$, with $^*F^{\mu\nu}=\frac12 \epsilon^{\mu\nu\rho\sigma}
F_{\rho \sigma}$ being the dual of $F_{\mu\nu}$. In a Lorentz frame
one has that the action (\ref{5.7}) is given by
\begin{equation}
{\cal L} = \sqrt{ 1 + ( {\bf B}^2 - {\bf E}^2 ) - ({\bf E} \cdot {\bf B})^2}.
	    \label{5.19}
\end{equation}
For the restricted set of fields the local equations ${^*{F^{\alpha\beta}}}_{,\alpha}$ 
yield the same two first equations as in (\ref{5.1}). 
The dynamical equation arise from varying the action (\ref{5.7}) with respect 
to $A_\alpha$ giving
\begin{equation}
{\cal L}^2 {F^{\alpha\beta}}_{,\alpha} - \lambda_\alpha F^{\alpha\beta}
+\rho_\alpha {^*{F^{\alpha\beta}}} = 0  \, ,
	    \label{5.20}
\end{equation}
where $\lambda_\alpha \equiv P_{,\alpha}-SS_{,\alpha}$
and $\rho_\alpha \equiv S\lambda_\alpha-L^2S_{,\alpha}$. If we develop 
(\ref{5.9}) for $\beta=x,y$ we obtain the following  equations,
\begin{eqnarray}
& 
 \left[1+{B_y}^2-{E_y}^2-{B_x}^2(B^2-E^2)-(E\cdot B)^2\right]\partial_0E_x
			 &\nonumber\\ & 
+\left[1+{B_x}^2-{E_x}^2+{E_y}^2(B^2-E^2)-(E\cdot B)^2\right]\partial_zB_y
			 &\nonumber\\ &
+\left[ (E_xB_y + E_yB_x)+(E\cdot B)(E_xE_y+B_xB_y)+(B^2-E^2)E_yB_x\right]
		    \partial_zE_x    &\nonumber\\ &
+\left[-(E_xB_y + E_yB_x)+(E\cdot B)(E_xE_y+B_xB_y)-(B^2-E^2)E_yB_x\right]
		    \partial_0B_y    &\nonumber\\ &
+\left[-2E_xB_x          +(E\cdot B)({E_x}^2+{B_x}^2)-(B^2-E^2)E_xB_x\right]
		    \partial_0B_x    &\nonumber\\ &
+\left[(E_xE_y - B_xB_y)+(E\cdot B)(E_xB_y-E_yB_x)+(B^2-E^2)E_xE_y\right]
		    \partial_zB_x    &\nonumber\\ &
+\left[(E_xE_y - B_xB_y)+(E\cdot B)(E_xB_y-E_yB_x)-(B^2-E^2)B_xB_y\right]
		    \partial_0E_y    &\nonumber\\ &
+\left[2E_yB_y          +(E\cdot B)({E_y}^2+{B_y}^2)+(B^2-E^2)E_yB_y\right]
		    \partial_zE_y =0    \, ,
	    \label{5.21}
\end{eqnarray}
and, 
\begin{eqnarray}
& 
 \left[1+{B_x}^2-{E_x}^2-{B_y}^2(B^2-E^2)-(E\cdot B)^2\right]\partial_0E_y
			 &\nonumber\\ & 
-\left[1+{B_y}^2-{E_y}^2+{E_x}^2(B^2-E^2)-(E\cdot B)^2\right]\partial_zB_x
			 &\nonumber\\ &
+\left[-(E_xB_y + E_yB_x)+(E\cdot B)(E_xE_y+B_xB_y)-(B^2-E^2)E_xB_y\right]
		    \partial_0B_x    &\nonumber\\ &
+\left[-(E_xE_y - B_xB_y)+(E\cdot B)(-E_xB_y+E_yB_x)-(B^2-E^2)B_xB_y\right]
		    \partial_0E_x    &\nonumber\\ &
+\left[-2E_yB_y          +(E\cdot B)({E_y}^2+{B_y}^2)-(B^2-E^2)E_yB_y\right]
		    \partial_0B_y    &\nonumber\\ &
+\left[-(E_xE_y - B_xB_y)+(E\cdot B)(E_xB_y-E_yB_x)-(B^2-E^2)E_xE_y\right]
		    \partial_zB_y    &\nonumber\\ &
-\left[(E_xB_y + E_yB_x)+(E\cdot B)(E_xE_y+B_xB_y)+(B^2-E^2)E_xB_y\right]
		    \partial_zE_y    &\nonumber\\ &
-\left[2E_xB_x          +(E\cdot B)({E_x}^2+{B_x}^2)+(B^2-E^2)E_xB_x\right]
		    \partial_zE_x  =0  \, .
	    \label{5.22}
\end{eqnarray}
Then using the column-vector ${\bf u}=(E_x,E_y,B_x,B_y)$ one finds that
the characteristic equation $\rm{det}(A^\mu\Sigma_\mu)=0$ is a quartic 
equation in $\Sigma_0$ and in $\Sigma_z$. Expanding this equation and using 
Boillat's results one finds that the characteristic equation is
\begin{equation}
\left[(P+1) g^{\alpha\beta} + \tau^{\alpha\beta}\right]
\left[(P-1) g^{\gamma\delta} + \tau^{\gamma\delta}\right]
\Sigma_\alpha\Sigma_\beta\Sigma_\gamma\Sigma_\delta=0 \, ,
	    \label{5.23}
\end{equation}
where
$\tau^{\alpha\beta}=P g^{\alpha\beta}-F^{\alpha\rho}{F^\beta}_\rho$ is the 
Maxwellian energy-momentum tensor. We see that in principle there is 
birefringence, i.e four solutions (two pairs of advanced and retarded waves) 
for the Born-Infeld theory. However, Boillat \cite{Boillat} shows that if 
the {\it weak energy condition} is obeyed then the only valid characteristic 
equation is given by
\begin{equation}
\left[(P+1) g^{\alpha\beta} + \tau^{\alpha\beta}\right]
\Sigma_\alpha\Sigma_\beta=0 \, .
	    \label{5.24}
\end{equation}
\vskip 0.4cm
{\tbf 5.4 Born-Infeld theory in Kaluza-Klein space}
\vskip 0.3cm
\indent
In five Kaluza-Klein dimensions one has to evaluate the determinant:
\begin{eqnarray}
&{\rm det} \, \pmatrix{1 & F^0_{\ \ x} & F^0_{\ \ y} & F^0_{\ \ z} & 
\partial^0 \phi 
\cr F^x_{\ \ 0} & 1 & F^x_{\ \ y} & F^x_{ \ \ z} &
\partial^x \phi 
\cr F^y_{\ \ 0} & F^y_{\ \ x} & 1 & F^y_{ \ \ z} &
\partial^y \phi 
\cr F^z_{\ \ 0} & F^z_{\ \ x} & F^z_{\ \ y} & 1 & 
\partial^z \phi 
\cr \partial_0 \phi & \partial_x \phi & \partial_y \phi & 
\partial_z \phi & 1} 
= &\nonumber\\&
=1 + \frac14 F_{\mu\nu}F^{\mu\nu}
-\frac{1}{16} {^*{F_{\mu\nu}}} F^{\mu\nu}
- \phi_{,\mu} \phi^{,\mu}
- \phi_{,\mu}\phi_{,\nu} (\frac12
F_{\alpha\beta}F^{\alpha\beta}g^{\mu\nu}-F^{\mu\alpha}{F^\nu}_\alpha)  \, .
	    \label{5.25}
\end{eqnarray}
Thus
\begin{equation}
{\cal L}=\sqrt{1+2P-S^2-2\Phi-2I}
\, .
	    \label{5.26}
\end{equation}
As before, 
$P=\frac14 F_{\mu\nu}F^{\mu\nu}$,  
$S=\frac14 {^*{F_{\mu\nu}}} F^{\mu\nu}$, 
$\Phi=\frac12 \phi_{,\mu}\phi^{,\mu}$, and \hfill\break
$I = \frac12 \phi_{,\mu}\phi_{,\nu}  (\frac12
F_{\alpha\beta}F^{\alpha\beta}g^{\mu\nu}-F^{\mu\alpha}{F^\nu}_\alpha)$. 
\newline
\indent
For the full case one encounters huge and intractable equations. In order to 
simplify the problem we suppose a constant electromagnetic background 
$F_{\mu\nu}={\rm a \, constant}$ tensor. In this case the Lagrangian 
reduces to
\begin{equation}
{\cal L}=\sqrt{1+2 \vartheta} \, ,
	    \label{5.27}
\end{equation}
where $\vartheta=\frac12 {\Theta}^{\mu\nu}\phi_{,\mu}\phi_{,\nu}$ and 
${\Theta}^{\mu\nu}$ is also a constant and symmetric tensor given by
\begin{equation}
{\Theta}^{\mu \nu} = \frac{F^{\mu \alpha} \, {F^{\nu}}_{\alpha} - (1+2P) 
g^{\mu \nu} }{\sqrt{1 + 2P-S^2}} \label{5.28}
\end{equation}
Then the equation of motion for $\phi$ is
\begin{equation}
(1+2 \vartheta) {\Theta}^{\alpha\beta}\phi_{,\beta,\alpha}
-{\Theta}^{\mu\nu}\phi_{,\mu} \phi_{\alpha,\nu}
{\Theta}^{\alpha \beta} \phi_{,\beta}=0 \, .
	    \label{5.29}
\end{equation}
We suppose now that the scalar field $\phi$ propagates in the $(t,z)$
directions with $\psi=\partial_0\phi$ and $\chi=\partial_z\phi$.  We
are also assuming, as before, $E_z=0$ and $B_z=0$.  In addition, by
choosing appropriately the directions of the fields $\bf E$ and $\bf B$
one can study a situation in which the relevant conponents of
$\Theta^{\mu\nu}$ are $\Theta^{00}$ and $\Theta^{zz}$ alone, with 
$\Theta^{0z}=0$. This can be achieved in three distinct situations: (i) 
$E_x=E_y=0$, (ii) $E_x=B_x=0$, or (iii) $E_y=B_y=0$. Then, $\Theta^{00}=
- \frac{1+{\bf B}^2}{\sqrt{1+2P-S^2}}$, and $\Theta^{zz}=
\frac{1-{\bf E}^2}{\sqrt{1+2P-S^2}}$ and $\Theta^{xx}$, $\Theta^{yy}$ 
are non-zero but not relevant to the problem.
To simplify the notation we write $\Theta^{00}=l$ and $\Theta^{zz}=m$. 
With these assumptions the equation of motion takes the form 
\begin{equation}
-l \, \partial_0\psi + m\partial_z\chi - l m \, \psi(\psi\partial_z\chi-
\chi\partial_z\psi)
-l m \, \chi(\chi\partial_0\psi-\psi\partial_0\chi)
\, .
			\label{5.30}
\end{equation}
Taking into account the internal equation $\partial_0\chi-\partial_z\psi=0$,  
the matrix ${\cal A^\mu}\Sigma_\mu$ is
\begin{equation}
{\cal A}^\mu\Sigma_\mu=
\pmatrix{
 -\Sigma_z & \Sigma_0
\cr 
-l \, (1+m\chi^2) \, \Sigma_0 + l m \, \psi\chi \, \Sigma_z 
& l m \, \psi \chi \, \Sigma_0 + m \,(1-l \, \psi^2) \, \Sigma_z}\, .
			\label{5.31}
\end{equation}
Its determinant yields the following characteristic equation
\begin{equation}
l \, {\Sigma_0}^2 - m \, {\Sigma_z}^2 + l m \, (\chi^2 \,{\Sigma_0}^2
- 2 \psi \chi \, \Sigma_0 \, \Sigma_z+ \psi^2 \, {\Sigma_z}^2)=0 \, . 
			\label{5.32}
\end{equation}
Equation (\ref{5.32}) can be written in a covariant form 
\begin{equation}
-\Theta^{\mu\nu} \, \Sigma_\mu \, \Sigma_\nu + (\Theta^{\alpha\mu} \,
\Theta^{\beta\nu}\phi_{,\alpha}\phi_{,\beta}- 2 \vartheta \, \Theta^{\mu\nu} ) 
\, \Sigma_\mu \, \Sigma_\nu =0 \, . 
			\label{5.33}
\end{equation}
In vacuum $\Theta^{\mu\nu}=-g^{\mu\nu}$ and (\ref{5.33}) reduces to the 
propagation of a scalar Born-Infeld field $g^{\mu\nu}\, \Sigma_\mu\Sigma_\nu 
+ (\phi^{,\mu}\phi^{,\nu}-g^{\mu\nu} \, \phi_{,\rho}\phi^{,\rho}) \,
\Sigma_\mu \, \Sigma_\nu=0$. 
\newline
\indent
In order to find $k(\omega)$ we put the ansatz $\Sigma=a{\rm e}^{i(\omega t -kz)}$
into equation (\ref{5.32})  yielding
\begin{equation}
k=\omega \, \sqrt{\frac{l}{m}} \,
\frac{\sqrt{1-l \, \psi^2+m \, \chi^2}- \sqrt{lm} \, \psi\chi}{1-l \,\psi^2} 
\, . 
			\label{5.34}
\end{equation}
One can study various limits. In particular,  an interesting case occurs 
when $\psi^2 \sim \chi^2 \sim \psi \chi \sim 0$ and $E^2\sim B^2<<1$. 
Then the velocity of propagation of the wave is $v=\frac{\omega}{k}=
\sqrt{\frac{l}{m}}=1-\epsilon$ with $\epsilon<<1$. In this example we recover
the propagation of a linear scalar field in smooth electromagnetic background. 
\vskip 0.4cm
\indent
{\tbf 6. Conclusion}
\vskip 0.3cm
\indent
In the present article we have shown how various multi-dimensional 
generalizations of the Born-Infeld theory lead to complicated interactions 
between the electromagnetic and dilaton fields. The exceptional character
of this theory is lost as soon as the dilaton field appears, induced
by the extra Kaluza-Klein dimension. Only in very low dimensions the
birefringence is not observed; in four space-time dimensions it appears
in all the generalizations of the Born-Infeld theory. 
\newline
\indent
Considering the Born-Infeld theory as derived from an effective Lagrangian
of a multi-dimensional string Lagrangian, we come to the conclusion that 
the features that were particularly interesting in classical version are
lost due to the emergence of highly non-linear couplings between all the
fields present in the theory.

\end{document}